\documentclass[%
 reprint,
 amsmath,amssymb,amsfonts
 aps,
prb,twocolumn
]{revtex4-2}
\usepackage[dvipdfmx]{graphicx}
\usepackage{graphicx}
\usepackage{bm}
\usepackage{txfonts}
\usepackage{color}
\usepackage{bm}
\usepackage[version=3]{mhchem}
\usepackage{mathtools}
\usepackage{newtxtext}
\usepackage[varg]{newtxmath}
\usepackage[normalem]{ulem}
\usepackage{lipsum}
\usepackage{here}
\usepackage{braket}
\usepackage{comment}
\usepackage[final]{hyperref}
\usepackage{txfonts}
\usepackage{color}
\usepackage[dvipdfmx]{graphicx}
\usepackage{here}
\usepackage{graphicx}
\usepackage{bm}
\usepackage{booktabs}
\usepackage{multirow}
\usepackage[mathlines]{lineno}

\usepackage{bm}
\usepackage{txfonts}
\usepackage{color}
\usepackage{color}
\usepackage{here}
\usepackage{dcolumn}
\usepackage{bm}
\usepackage[version=3]{mhchem}
\usepackage{mathtools}
\usepackage{newtxtext}
\usepackage[varg]{newtxmath}
\usepackage[normalem]{ulem}
\usepackage{lipsum}
\usepackage{braket}
\usepackage{comment}
\usepackage[final]{hyperref}

\begin{document}
\title{First-principles study of exchange stiffness constant of half{-}metallic Heusler alloys $\textrm{Co}_2\textrm{MnZ}$ (Z= Si, Al) at finite temperatures: {Spin fluctuation}{-}induced effective half metallicity  }


\author{Shogo Yamashita}\email{shogo.yamashita.q1@dc.tohoku.ac.jp} 
 \author {Akimasa Sakuma} 
\affiliation{Department of Applied Physics, Tohoku University, Sendai 980-8579, Japan}

\author{Mikihiko Oogane\thanks{Job elsehere} }
\affiliation{Department of Applied Physics, Tohoku University, Sendai 980-8579, Japan
}



\date{\today}
\begin{abstract} {We performed first-principles calculations at finite temperatures to investigate {the} temperature dependence of the magnetic properties, such as exchange stiffness constants and Curie temperatures, of $\textrm{Co}_2\textrm{MnZ}$ (Z= Si, Al) assuming $\textrm{L}2_{1}$ and $\textrm{B}2$ structures. In $\textrm{L}2_{1}$ structures, we confirmed {a} relatively high Curie temperature for $\textrm{Co}_2\textrm{MnAl}$, compatible {with} that of $\textrm{Co}_2\textrm{MnSi}${;} however, its exchange stiffness constant and single site magnetic excitation energy at zero temperature are much smaller than {those} of $\textrm{Co}_2\textrm{MnSi}$. This might indicate that the Curie temperature of itinerant magnets cannot be determined by the exchange interaction at zero temperature. We also investigated {the} temperature dependence of {the} exchange stiffness constants of both alloys, and we found robustness {in the} temperature dependence of {the} exchange stiffness constant for $\textrm{Co}_2\textrm{MnAl}$, assuming {an} $\textrm{L}2_{1}$ structure. This might lead {to a} high Curie temperature, contrary to {the} small exchange stiffness constant. Finally, we examined the temperature dependence of the electronic structure to investigate the origin of {the} behavior of {the} exchange stiffness constant at finite temperatures. We confirmed that {the} spin polarization at chemical potential effectively increases with {an increasing}  temperature due to {the altered} electronic structure induced by the spin disorder. This might contribute to the robustness of the exchange stiffness constant at finite temperatures. Our results might indicate that renomarization of {the} electronic structure due to spin disorder at finite temperature influences the exchange interactions of $\textrm{Co}_2\textrm{MnAl}$. } 
\end{abstract}
\maketitle
\section{Introduction}
Half-metallic Co-based full-Heusler alloys $\textrm{Co}_2\textrm{MnZ}$ (Z=Si, Al) are technologically important materials in the field of spintronics due to {their} high Curie temperatures, large magnetization, and theoretically predicted high spin polarization, {approximately} 100\% at the Fermi level\cite{Kubler1983, Galanakis2002,Galanakis2006}. 
A notable {application} example of $\textrm{Co}_2\textrm{MnSi(Al)}$ is {that} for spintronics devices based on magnetic tunnel junctions (MTJ), which exhibit a large tunnel magnetic resistance (TMR) effect. In fact, experimentally, {a} large TMR ratio has been achieved in MTJs where $\textrm{Co}_2\textrm{MnSi(Al)}$ {is} used for {the} electrodes, not only {at} low temperatures but also {at} room temperatures and half-metallic characters have been confirmed\cite{SakurabaAPL2006,Sakuraba2APL2006,Oogane2006,Sakuraba3APL2006,Tsunegi2008}. 
\ The TMR ratio usually decreases with {an increasing} temperature, and it is mainly considered that thermal spin fluctuations at the ferromagnet and barrier {interface} cause {this} reduction\cite{Tsunegi2009}. In addition, {it has also been theoretically} confirmed that {the} exchange interactions at {the} $\textrm{Co}_2\textrm{MnSi(Al)}$/MgO {interface} become weak compared {with those in} the of bulk, in particular, {the} exchange interaction at Co sites\cite{Sakuma2009}, and  {the} thermal spin fluctuations (non-collinear spin structure) at  {the} $\textrm{Co}_2\textrm{MnSi}$/MgO  {interface} reduce the TMR ratio\cite{Miura2011}.
Therefore, to improve the TMR ratio at finite-temperatures, it is important to enhance {the} exchange stiffness constant $A$ (or spin stiffness constant $D$) at the interface to suppress thermal spin fluctuations. The exchange stiffness constant is one of the indicators {that} show the strength of the exchange interactions in magnetic materials, namely, {the} robustness of the magnetic moments against thermal spin fluctuations. It is also {an} important parameter for micromagnetic simulations to calculate the dynamics of magnetization, as well as the magnetic anisotropy constant and Gilbert damping constant. \\
\ Apart from {the} interfaces, the temperature dependence of exchange stiffness constants (or thermal spin fluctuation) in $\textrm{Co}_2\textrm{MnSi(Al)}$ compounds may also contribute to {a secondary} reduction {in} the TMR ratio at finite temperatures. 
The exchange stiffness constants of $\textrm{Co}_2\textrm{MnSi(Al)}$ compounds have been measured in several experiments\cite{Ritchie2003,Rameev2006,Hamrle2009,Kubota2009,Umetsu2011}. Among them, Kubota et al.\cite{Kubota2009} reported a significantly smaller ({five} times) exchange stiffness constant of $\textrm{Co}_2\textrm{MnAl}$ compared with that of $\textrm{Co}_2\textrm{MnSi}$, which {were} measured by them or Hamrle et al.\cite{Hamrle2009} at room temperature with Brillouin light scattering spectroscopy. 
{Meanwhile}, {the} experimental results of the spin stiffness constants measured by Umetsu et al.\cite{Umetsu2011} from the temperature dependence of  magnetization at low temperature are not so different between $\textrm{Co}_2\textrm{MnSi}$ and $\textrm{Co}_2\textrm{MnAl}$.
Therefore, to explain this experimental difference, {a} comprehensive theoretical understanding of {the} exchange stiffness constants, not only {at} low-temperatures but also {at} room temperature or {the entire} temperature range, may be required. \\
\ Theoretically, {the} temperature dependence of the exchange stiffness constants is described by the magnon magnon scattering theory based on the localized Heisenberg Hamiltonian as proposed by Dyson\cite{Dyson1956}. However, nowadays, $3d$ electrons in the metallic magnetic systems are regarded as itinerant electrons. They have contradictory dual properties {including} hybridization effects with other orbitals and large local magnetic moments, which are non-integer values\cite{Moriyabook,KublerBook}. In fact, it {has been} theoretically shown that the gaps in the minority spin states of $\textrm{Co}_2\textrm{MnSi(Al)}$ originate {from} the hybridization of  $d$ orbitals of Co-Co atoms, and the hybridization of Mn-Co also {plays an} important role to determine its electronic structure\cite{Galanakis2002,Galanakis2006}. Moreover, it {has been} shown that {the} magnetic moments of Co atoms are less localized compared { with those} of Mn atoms {using} x-ray absorption and x-ray magnetic circular dichroism techniques\cite{Telling2006,Telling2008,Fecher2014}.
In this sense, $\textrm{Co}_2\textrm{MnSi(Al)}$ {compounds} may be typical examples {where} $3d$ electrons have dual itinerant and localized properties. Therefore, the localized spin models, such as the Heisenberg model, which are appropriate for magnetic insulators or $4f$ electron systems, may not be appropriate {approaches} for these alloys to capture magnetic properties, not only { at} zero temperature but also { at} finite temperatures.
Based on { the} above mentioned concept, in previous works\cite{Sakuma2024,Yamashita2024-2}, we established  { a} theory to describe  { the} temperature dependence of the exchange stiffness constants based on the functional integral method for itinerant electron systems\cite{Cyrot,Hubbard1,Hubbard2,Hasegawa1,Hasegawa2,Hasegawa3,Moriyabook}, combined with the disordered local moment (DLM) picture based on  the coherent potential approximation (CPA). The DLM-CPA theory was originally proposed in the first-principles calculation based on the Korringa-Kohn-Rostoker method\cite{Oguchi,Pindor,Staunton3,Gyorrfy,StauntonPRL1992,StauntonPRL,StauntonPRB}, and we reformulated it in the tight-binding linear muffin-tin orbital (TB-LMTO) method in our previous works\cite{Hiramatsu1,Sakuma2022,Yamashita2022,Hiramatsu2023,Yamashita2023,Sakuma2024,Yamashita2024-2}.   
Therefore, by using our developed theory, we can investigate the exchange stiffness constant of the Heusler alloys at finite temperatures based on the electronic structures, which include both itinerant and localized characters of $d$ electrons.
In this study we investigated the exchange stiffness constants of $\textrm{Co}_2\textrm{MnSi(Al)}$  { with} $\textrm{B}2$ or $\textrm{L}2_{1}$ structures, not only  { at} zero temperature but also  { at} finite temperatures,  and found unique behavior for $\textrm{Co}_2\textrm{MnAl}$ with $\textrm{L}2_{1}$ structure at finite temperatures.
We also examined the effect of spin fluctuation (spin disorder) at finite-temperatures for the electronic structures and exchange interactions of $\textrm{Co}_2\textrm{MnSi(Al)}$ and discussed the relationship between them.
\section{Theoretical method}
We use the TB-LMTO method\cite{Andersen,Turek,Kudrnovsky,Sakuma2000,Skriver} combined with the atomic sphere approximation (ASA) to construct  { a} first-principles functional integral theory.
The functional integral to evaluate the partition function $Z$ and free energy $F$ are given as follows:
\begin{align}
Z=\int \left(\prod_{i} \textrm{d} {\bf{e}}_{i} \right) \textrm{exp} \left({-\Omega(T,\lbrace{\bf {e}}\rbrace)/k_{\textrm{B}}T}\right),
\end{align}
\begin{align}
F=\langle \Omega(T,\lbrace{\bf {e}}\rbrace)\rangle_{\omega(T,\lbrace{\bf {e}}\rbrace)}+k_{\textrm{B}}T \langle \textrm{ln} \ \omega(T,\lbrace{\bf {e}}\rbrace)\rangle_{\omega(T,\lbrace{\bf {e}}\rbrace)}+\mu N_{\textrm{e}},
\label{FreeE}
\end{align}
where 
\begin{align}
\langle \cdots \rangle_{\omega(T,\lbrace{\bf {e}}\rbrace)}=\int \left( \prod_{i} \textrm{d} {\bf {e}}_{i} \right) \omega(T,\lbrace{\bf {e}}\rbrace) \cdots.
\end{align}
Here, $\mu$ and $N_{\textrm{e}}$ are chemical potential and electron number of the system, respectively.
The thermodynamic potential $\Omega$ can be decomposed as follows:
\begin{align}
\Omega=\Omega_{\textrm{SP}}+\Omega_{\textrm{DC}},
\end{align}
where $\Omega_{\textrm{SP}}$ and $\Omega_{\textrm{DC}}$ are  { the} thermodynamic potentials of  { the} single-particle and double counting part.
{The thermodynamic potentials are evaluated within  { an} adiabatic approximation because of  low energy scale (slow motion) of spin fluctuations compared to  { the} energy scale of hopping of electrons.
Here, to simplify the theory, we are based on the magnetic force theorem to evaluate  { the} magnetic excitation energy with respect to its directions keeping its lengths. This means that only spin-transverse fluctuations are included and the longitudinal fluctuations are ignored in our formalism}.  In addition, we use the local spin density approximation (LSDA) as the exchange correlation potential to calculate  { the} thermodynamic potential of  { the} single-particle part. Therefore,  { the} distribution of { the} spin configuration $ \omega(T,\lbrace{\bf {e}}\rbrace)$ at temperature $T$ is given as follows:
\begin{align}
 \omega(T,\lbrace{\bf {e}}\rbrace)\sim\textrm{exp} \left(-\Omega_{\textrm{SP}}(T,\lbrace{\bf {e}}\rbrace)/k_{\textrm{B}}T \right)/Z_{\textrm{SP}},
\end{align}
\begin{align}
\Omega_{\textrm{SP}}(T,\lbrace{\bf {e}}\rbrace)=\frac{1}{\pi}\int \textrm{d} \epsilon f(\epsilon,T,\mu)\int_{-\infty}^{\epsilon}\textrm{d} E {\textrm{Im}}{\textrm{Tr}}\ G(z;{\lbrace{\bf {e}}\rbrace}),
\end{align}
where $z=E+i\delta$. $\delta$ is fixed to 2 mRyd. { Owing to the force theorem, we  dropped the double counting term to evaluate { the} spin configuration $ \omega(T,\lbrace{\bf {e}}\rbrace)$. }
Here, the Green function of the system $G(z;{\lbrace{\bf {e}}\rbrace})$ is introduced as follows:
 \begin{align}
&G_{ij}(z;{\lbrace{\bf {e}}\rbrace})=(z-H_{{\textrm{TB-LMTO}}}(\lbrace{\bf {e}}\rbrace))_{ij}^{-1} \nonumber \\
&=\tilde \lambda_{i}^{\alpha}(z;{\bf e}_{i})\delta_{ij}+\tilde\mu^{\alpha}_{i} (z;{\bf e}_{i}) g_{ij}^{\alpha} (z;{\lbrace{\bf {e}}\rbrace}) (\tilde \mu_{j}^{\alpha}(z;{\bf e}_{j}))^{\dag}, 
\end{align}
\begin{align}
&H_{{\textrm{TB-LMTO}}}(\lbrace{\bf {e}}\rbrace) \nonumber \\
&=\tilde C(\lbrace{\bf {e}}\rbrace)+\tilde \Delta^{\frac{1}{2}}(\lbrace{\bf {e}}\rbrace) S(1-\tilde \gamma(\lbrace{\bf {e}}\rbrace) S)^{-1} \tilde \Delta^{\frac{1}{2}}(\lbrace{\bf {e}}\rbrace),
\end{align}
where $\tilde A(\lbrace{\bf {e}}\rbrace)$ is given as $U^{\dag}(\lbrace{\bf {e}}\rbrace)AU(\lbrace{\bf {e}}\rbrace)$ and $C$, $\gamma$, $\Delta$ are potential parameters in the TB-LMTO method. $S$ is a bare structure constant. 
$\lambda^{\alpha}_{i}$ and $\mu^{\alpha}_{i}$ are given as follows:
\begin{align}
\lambda^{\alpha}_{i}(z)=\Delta_{i}^{-\frac{1}{2}} \left(1+(\gamma_{i}-\alpha)P_{i}^{\gamma}(z)\right)\Delta_{i}^{-\frac{1}{2}},
\end{align}
\begin{align}
\mu^{\alpha}_{i}(z)=\Delta_{i}^{-\frac{1}{2}} \left(P_{i}^{\gamma}(z)\right)^{-1} P_{i}^{\alpha}(z),
\end{align}
\begin{align}
  (\mu_{i}^{\alpha}(z))^{\dag}=P_{i}^{\alpha}(z) \left(P_{i}^{\gamma}(z)\right)^{-1}  \Delta_{i}^{-\frac{1}{2}},
\end{align}
\begin{align}
P_{i}^{\gamma}(z)= \Delta_{i}^{-\frac{1}{2}}(z-C_{i}) \Delta_{i}^{-\frac{1}{2}} ,
\end{align}
\begin{align}
 P_{i}^{\alpha}(z)= P_{i}^{\gamma}(z) \left[1-(\alpha-\gamma_{i})P_{i}^{\gamma}(z)\right]^{-1},
\end{align}
\begin{align}
 g^{\alpha}_{ij}(z;{\lbrace{\bf {e}}\rbrace})=\left[(\tilde P^{\alpha}(z;{\lbrace{\bf {e}}\rbrace})-S^{\alpha})^{-1}\right]_{ij},
\end{align}
where $\alpha$ is a site-independent matrix of maximum localized representation,  { which is} summarized in several references.
Here, $S^{\alpha}$ is given as $S(1-\alpha S)^{-1}$.
 { In practice}, we cannot treat various patterns of spin configuration $\lbrace{\bf {e}}\rbrace$ in the functional integral. Therefore, we adopt the CPA and single-site approximation to determine $\omega(T,\lbrace{\bf {e}}\rbrace)$. Here, we decompose $\Omega_{\textrm{SP}}(T,\lbrace{\bf {e}}\rbrace)$ and $\omega(T,\lbrace{\bf {e}}\rbrace)$ as follows:
 \begin{align}
 \Omega_{\textrm{SP}}(T,\lbrace{\bf {e}}\rbrace)= \Omega^{0}_{\textrm{SP}}+\sum_{i} \Delta \Omega_{\textrm{SP}}(T,{\bf {e}}_{i}),
 \label{SingleOmega}
\end{align}
 \begin{align}
\omega(T,\lbrace{\bf {e}}\rbrace)=\prod_{i}\omega_{i}(T,{\bf {e}}_{i}).
\end{align}
The first term in Eq. \eqref{SingleOmega} is independent on spin configuration. 
Therefore,  { the} decomposed distribution of spin configuration $\omega_{i}(T,{\bf {e}}_{i})$ is calculated by using only  { the} second term in  { Eq.} \eqref{SingleOmega} as follows:
\begin{align}
&\omega_{i}(T,{\bf {e}}_{i})\nonumber \\ &=\textrm{exp} \left(-\Delta\Omega_{\textrm{SP}}(T,{\bf {e}}_{i})/k_{\textrm{B}}T \right)/\int \textrm{d} {\bf {e}}_{i} \textrm{exp} \left(-\Delta\Omega_{\textrm{SP}}(T,{\bf {e}}_{i})/k_{\textrm{B}}T\right).
\end{align}
Here, we can obtain $\Delta\Omega_{\textrm{SP}}(T,{\bf {e}}_{i})$ from  { the} following equations:
 \begin{align}
&\Delta\Omega_{\textrm{SP}}(T,{\bf {e}}_{i})\nonumber \\
&=-\frac{1}{\pi} \int \textrm{d} \epsilon f(\epsilon,T,\mu)) \textrm{Im} [ \textrm{Tr}_{L\sigma} \ \textrm{log}  \lambda_{i}^{\alpha}(z) \nonumber \\ 
&-\textrm{Tr}_{L\sigma} \ \textrm{log}(1+\Delta P_{i}(z;{\bf {e}}_{i}) \bar g_{ii}^{\alpha}(z))],
\label{DOsp}
\end{align}
 \begin{align}
\Delta P_{i}(z;{\bf {e}}_{i}) =\tilde P_{i}^{\alpha}(z;{\bf {e}}_{i})-\bar P_{i}(z),
\label{DP}
\end{align}
 \begin{align}
\bar g^{\alpha}(z)=\left(\bar P(z)-S^{\alpha}\right)^{-1},
\label{CG}
\end{align}
where $\bar g^{\alpha}(z)$ and $\bar P(z)$ are the coherent Green function and coherent potential function.
 Here, we used  { the} fact that $\textrm{Tr}_{L\sigma}\textrm{ln}\  \tilde\lambda_{i}^{\alpha}(z;{\bf {e}}_{i})=\textrm{Tr}_{L\sigma}\textrm{ln}\ \lambda_{i}^{\alpha}(z)$, and thus  { the} first term of Eq. \eqref{DOsp} becomes independent  { of the} spin configuration. $\bar g^{\alpha}(z)$ and $\bar P(z)$ are obtained from  { the} following CPA condition together with Eq. \eqref{DP} and \eqref{CG}  in  { a} self-consistent manner:
\begin{align}
{\int \textrm{d} {{\bf e}}_{i} \ \omega_{i}(T,{\bf{e}}_{i})  \left[1+ \Delta P_{i}(z;{\bf e}_{i})\bar g^{\alpha}_{ii}(z)\right]^{-1} \Delta P_{i}(z;{\bf e}_{i})=0.}
\label{CPAcon}
\end{align}
Once we obtain  { the} decomposed distribution of spin configuration $\omega_{i}(T,{\bf {e}}_{i})$, we can calculate { the} thermodynamic potential of { the} single particle part as follows:
\begin{align}
\langle \Omega_{\textrm{SP}}\rangle_{\lbrace{\omega_{i}(T,{\bf {e}}_{i})}\rbrace}&= \sum_{i}\int \textrm{d} {\bf e}_{i} \  \omega_{i}(T,{\bf{e}}_{i}) \  \Omega_{\textrm{SP}}(T,{\bf e}_{i}),
\label{ENESOC}
\end{align}
\begin{align}
& \Omega_{\textrm{SP}}(T,{\bf e}_{i})=-\frac{1}{\pi} \textrm{Im}\textrm{Tr}_{L\sigma} \int _{-\infty}^{\infty}\textrm{d}  \epsilon \ \epsilon f(\epsilon,T,\mu) G_{ii}(\epsilon+i\delta,{\bf e}_{i}) \nonumber \\ &-\frac{k_{\textrm{B}}T}{\pi}\textrm{Im}\textrm{Tr}_{L\sigma}\int _{-\infty}^{\infty}\textrm{d}  \epsilon \ G_{ii}(\epsilon+i\delta,{\bf e}_{i}) \nonumber \\ &\times \left(  f(\epsilon,T,\mu) \textrm{log} f(\epsilon,T,\mu)+(1- f(\epsilon,T,\mu)) \textrm{log} (1- f(\epsilon,T,\mu))  \right),
\label{Omega}
\end{align}
 \begin{align}
&G_{ii}(z,{\bf e}_{i})\nonumber \\ &=\tilde \lambda^{\alpha}_{i}(z,{\bf {e}}_{i})+\tilde \mu^{\alpha}_{i}(z,{\bf {e}}_{i})   \bar g^{\alpha}_{ii}(z) (1+\Delta P_{i}(z,{\bf {e}}_{i})\bar g^{\alpha}_{ii}(z))^{-1} \left( \tilde \mu^{\alpha}_{i}(z,{\bf {e}}_{i})\right)^\dag.
\label{Gii}
 \end{align} 
Hereafter, we introduce an external force  to express { the} spin-spiral magnetic structures labeled by $\vec q$ in the magnetically disordered environment at finite temperatures.
For this purpose, we use the generalized Bloch's theorem\cite{Sandratskii1986,Mryasov1992,Uhl1992,Kubler2006} and transform { the} bare structure constant as follows\cite{Yamashita2023APEX, Yamashita2024,Sakuma2024}:
 \begin{align}
 \tilde S_{ij}(\vec k,\vec q)= U_i \begin{pmatrix}
 S_{ij}(\vec k-\frac{\vec q}{2}) & 0 \\
0 & S_{ij}(\vec k+\frac{\vec q}{2}) \\
\end{pmatrix}U^{\dag}_j{,}
\label{SSP}
 \end{align}
 {where $i$ and $j$ stand for sites of { the} {sub-lattice in each} unit cell, and we redefine $U$ at site $i$ as follows:}
 \begin{align}
 U_{i}=\frac{1}{\sqrt{2}}
 \begin{pmatrix}
   e^{i(\frac{\vec q}{2}\cdot \vec r_{i})}& e^{-i(\frac{\vec q}{2}\cdot \vec r_{i})} \\
   -e^{i(\frac{\vec q}{2}\cdot \vec r_{i})}& e^{-i(\frac{\vec q}{2}\cdot \vec r_{i})}  \\
\end{pmatrix},
 \end{align}
 where $\vec r_{i}$ denotes {sub-lattice} positions in {each} unit cell. 
 Therefore, in the  spin-spiral magnetic state, the coherent Green function in Eq. \eqref{Gii} is replaced as follows:
 \begin{align}
 \bar g^{\alpha}_{ii}(z,\vec q)=\frac{1}{N}\sum_{\vec k} \left[(\bar P(z)-S^{\alpha}(\vec k, \vec q))^{-1}\right]_{ii},
 \label{spcgf}
 \end{align}
 where $N$ is a number of $k$-points in the Brillouin zone.
By using Eqs. \eqref{ENESOC} , \eqref{Omega}, \eqref{Gii}, and \eqref{spcgf}, we calculate { the} free energy difference at temperature $T$ as follows:
 \begin{align}
\Delta F(\vec q, T)\sim \langle \Omega_{\textrm{SP}}(\vec q)\rangle_{\lbrace{\omega_{i}(T,{\bf {e}}_{i})}\rbrace}-\langle \Omega_{\textrm{SP}}(\vec q=0)\rangle_{\lbrace{\omega_{i}(T,{\bf {e}}_{i})}\rbrace}.
\end{align}
It { can be} noted that ${\omega_{i}(T,{\bf {e}}_{i})}$ has $\vec q$ dependence; however, we focus on the vicinity of $\vec q=0$ to calculate { the} exchange stiffness constant. Therefore, we neglect { the} $\vec q$ dependence of  ${\omega_{i}(T,{\bf {e}}_{i})}$.
Here, we use { the} magnetic force theorem again and { do} not include the double counting term to calculate the free energy difference.
Once we { obtain the} free energy difference $\Delta F(\vec q, T)$, we can calculate the exchange stiffness constant at finite temperatures as follows:
\begin{align}
 A_{\eta}(T)=\frac{1}{2V} { {\left. \frac{d^2 \Delta F (\vec q, T)}{ d q_{\eta}^2} \right |_{q_{\eta}=0}.}}
\label{exA}
\end{align}
In this study, $\eta$ is fixed to the $z$-direction.
In particular, the spin stiffness constant at zero temperature $D$ is given as follows:
\begin{align}
 D_{\eta}=\frac{4V}{M}  A_{\eta}(0),
 \label{exD}
\end{align}
where $M$ and $V$ are { the} magnetization at zero temperature and a volume of the unit cell, respectively.
The details of the DLM-CPA method in the TB-LMTO method are also summarized in references\cite{Hiramatsu1,Sakuma2022,Yamashita2022,Hiramatsu2023,Yamashita2023,Sakuma2024,Yamashita2024-2}.
$20\times20\times20$ $k$-points are used for the calculations of exchange stiffness constants, and $30\times30\times30$ $k$-points are used for the calculation of { the} spin polarization ratio. The structure constants are expanded up to $d$-orbitals. { The lattice constants are set to 5.654 \AA \ and 5.756 \AA \ for $\textrm{Co}_2\textrm{MnSi}$ and $\textrm{Co}_2\textrm{MnAl}$ with $\textrm{L}2_{1}$ structure, respectively. For B2 structure, half values of them are adopted. The spin-orbit coupling is ignored in this study. }
\section{Results}
\subsection{Zero-temperature properties}
\begin{flushleft} 
\begin{figure}[h]
\begin{center}
\includegraphics[clip,width=8.5cm]{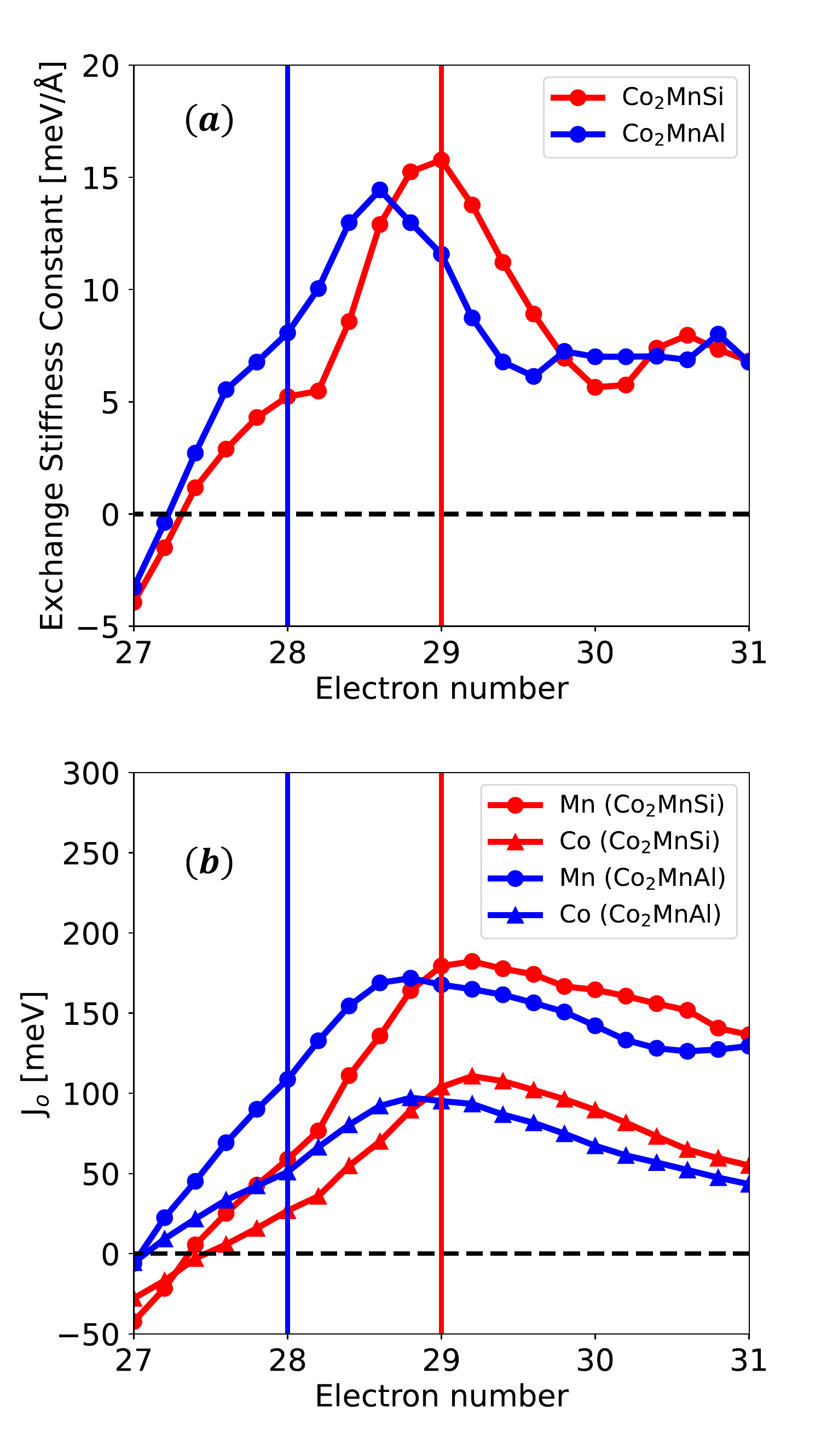}
\end{center}
\caption{Electron filling dependence of (a) exchange stiffness constant $A$ and (b) $J_{o}$ at Mn and Co sites of $\textrm{Co}_2\textrm{MnSi}$ (red lines) and $\textrm{Co}_2\textrm{MnAl}$ (blue lines) with $\textrm{L}2_{1}$ structure at zero temperature. { The} vertical lines correspond to { the} valence electron number of { the} corresponding alloy. }
\label{NdepAJ}
\end{figure} 
\end{flushleft}

\begin{flushleft} 
\begin{figure}[h]
\begin{center}
\includegraphics[clip,width=9cm]{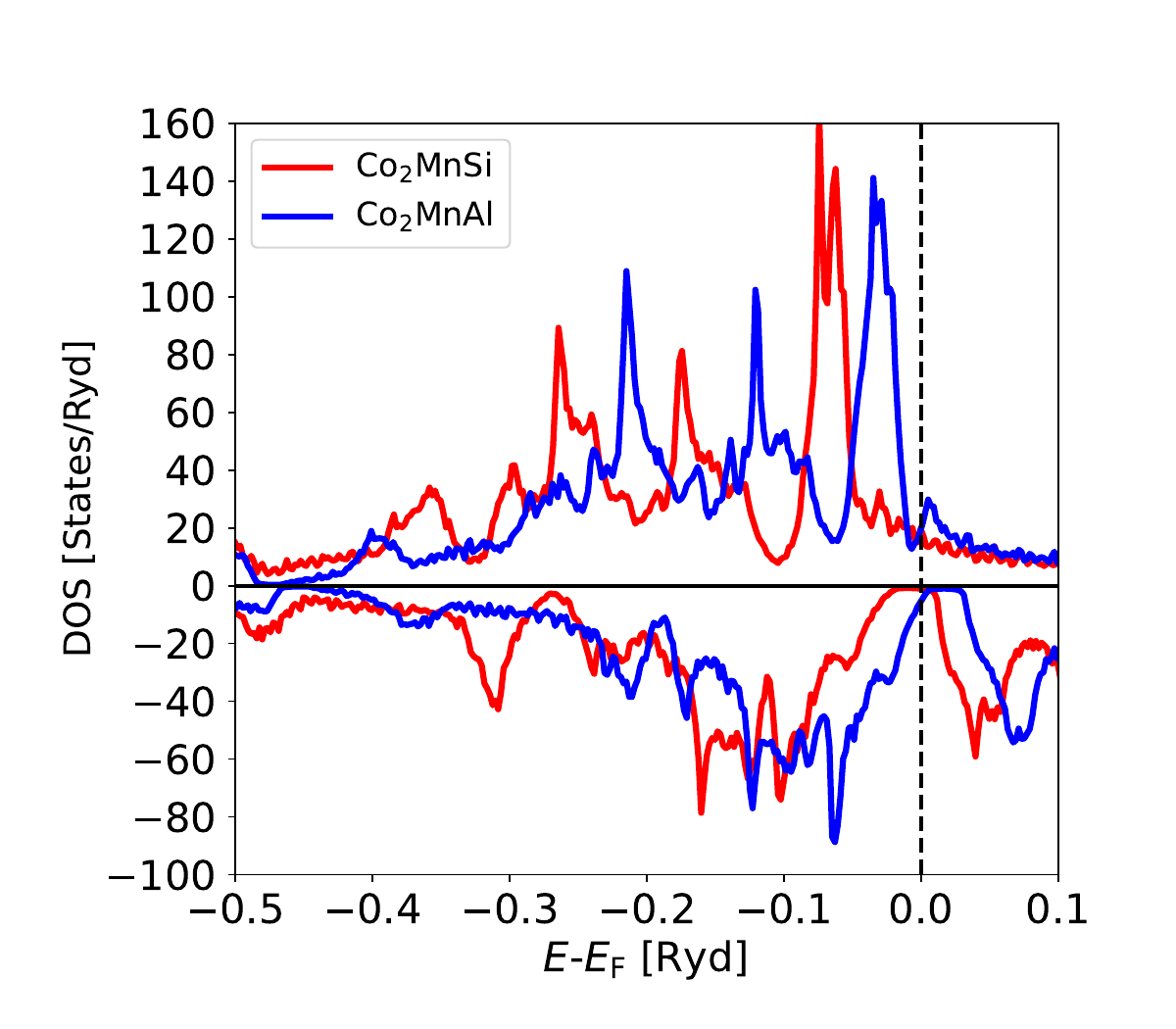}
\end{center}
\caption{Density of states of  $\textrm{Co}_2\textrm{MnSi}$ (Red) and $\textrm{Co}_2\textrm{MnAl}$ (Blue) with $\textrm{L}2_{1}$ structure at zero temperature.}
\label{DOS}
\end{figure} 
\end{flushleft}

We summarize calculated { the} magnetic properties of $\textrm{Co}_2\textrm{MnSi}$ and $\textrm{Co}_2\textrm{MnAl}$ at zero temperature in Table \ref{DATA}.
From this table, we can { observe} that the exchange stiffness constant of $\textrm{Co}_2\textrm{MnAl}$ with  $\textrm{L}2_{1}$ structure is almost half compared with that of  $\textrm{Co}_2\textrm{MnSi}$ with $\textrm{L}2_{1}$ structure.
However,  if we focus on the results for B2 structures, the exchange stiffness constants of $\textrm{Co}_2\textrm{MnSi}$ and $\textrm{Co}_2\textrm{MnAl}$ do not differ { significantly from} each other.  { The} calculated magnetic moments are also summarized in Table \ref{DATA}.
{ The} total magnetic moments of $\textrm{Co}_2\textrm{MnSi}$ and $\textrm{Co}_2\textrm{MnAl}$ with $\textrm{L}2_{1}$ structure are close to 5 $\mu_{\textrm{B}}$ and 4 $\mu_{\textrm{B}}$, respectively. Our calculated magnetic moments are in good agreement with { those in} previous studies\cite{Galanakis2002,Galanakis2006,Kubler2007,Nawa2020} with the LSDA. 
\begin{table*}
    \centering
    \caption{Calculated magnetic properties of $\textrm{Co}_2\textrm{MnSi}$ and $\textrm{Co}_2\textrm{MnAl}$ at zero temperature. }
    \begin{ruledtabular}
    \begin{tabular}{ccccccc}
        System &$A$ [meV/\AA] & $D$ [meV\ $\textrm{\AA}^2$] & $M_{\textrm{Co}}$ [$\mu_{\textrm{B}}$]   & $M_{\textrm{Mn}}$ [$\mu_{\textrm{B}}$] &  $M_{\textrm{Total}}$ [$\mu_{\textrm{B}}$] \\
    \hline  
            $\textrm{Co}_2\textrm{MnSi}$  $\textrm{L}2_{1}$    & 15.79 &  571.9   &    1.090    &    2.886         &    4.992     \\
        $\textrm{Co}_2\textrm{MnAl}$  $\textrm{L}2_{1}$     & 8.07   &   374.4    &  0.812     &     2.637      &     4.120     \\
   \hline   
       $\textrm{Co}_2\textrm{MnSi}$  B2    &    11.98    &  435.1 & 1.088     &     2.890        &   2.497 \\
      $\textrm{Co}_2\textrm{MnAl}$   B2     &   10.08    &  467.2  & 0.785      &    2.654       &    2.057     \\
   \end{tabular}
    \label{DATA}
    \end{ruledtabular}
\end{table*} \\
\ { T}o investigate the reason why the exchange stiffness constant of $\textrm{Co}_2\textrm{MnAl}$ with $\textrm{L}2_{1}$ structure is much smaller than that of $\textrm{Co}_2\textrm{MnSi}$, we examined the electron filling dependence of the exchange stiffness constant for both alloys.
In addition, we also examined the on-site magnetic excitation energy of the infinitesimal spin-rotation $J_{o}$ at site $o$ with the so called Lichtenstein-Katsnelson-Antropov-Gubanov (LKAG) formula\cite{Lichtenstein1987,Sakuma1999} as follows:
\begin{align}
J_{o}&=-\frac{1}{4\pi} \textrm{Im} \int_{-\infty}^{E_{\textrm{F}}} \textrm{d} \epsilon \  \textrm{Tr}_{L} \lbrace\Delta P^{\gamma}_{o} (\epsilon) \left(g^{\gamma}_{oo\uparrow\uparrow}(\epsilon)-g^{\gamma}_{oo\downarrow\downarrow}(\epsilon)\right) \nonumber \\
&+\Delta P^{\gamma}_{o}(\epsilon) g^{\gamma}_{oo\uparrow\uparrow}(\epsilon) \Delta P^{\gamma}_{o} (\epsilon)g^{\gamma}_{oo\downarrow\downarrow}(\epsilon)\rbrace,
\end{align} 
where we defined $\Delta P^{\gamma}_{o}(\epsilon)=P^{\gamma\uparrow}_{o}(\epsilon)-P^{\gamma\downarrow}_{o}(\epsilon)$.
 We show { the} calculated electron filling dependence of $A$ and $J_{o}$ for $\textrm{Co}_2\textrm{MnSi}$ and $\textrm{Co}_2\textrm{MnAl}$ with $\textrm{L}2_{1}$ structures in Fig. \ref{NdepAJ}.
Here, we assumed the rigid band model.  We changed { the} total electron number in the unit cell to shift the Fermi level and calculated $A$ and $J_{o}$ for each Fermi level. 
From Figs. \ref{NdepAJ} (a) and (b), we can { observe} that the behaviors of $A$ and $J_{o}$ are mostly { the} same { with} increasing electron filling.
In the case of $\textrm{Co}_2\textrm{MnSi}$, both $A$ and $J_{o}$ of Mn and Co atoms have almost maximum values { when the} electron number $N_{\textrm{e}}$ is 29.
{ Meanwhile}, in the case of $\textrm{Co}_2\textrm{MnAl}$, both $A$ and $J_{o}$ are far away from the maximum values at $N_{\textrm{e}}=28$.
 In addition, it is noted that $J_{o}$ { values} at Co sites are smaller than { those} of Mn sites in both cases. 
 This might indicate { that the} electronic states of Co sites are { more} easily affected by the thermal spin fluctuation than { those of} Mn sites. 
{ If we \textit{assume} that $A$ and $J_{o}$ at zero temperature determine the Curie temperatures and compare { the} empirical results, { which shows} that the Curie temperatures follow { an} almost linear behavior with respect to { the} magnetic moments (or number of valence electrons)\cite{Fecher2006}, our results are quite reasonable. We discuss { this point} deeply in { a} later section.}
At this stage, we might conclude that this difference of electron filling contributes to the significant difference { in} the exchange stiffness constants of $\textrm{Co}_2\textrm{MnSi}$ and $\textrm{Co}_2\textrm{MnAl}$ with $\textrm{L}2_{1}$ structure at zero temperature.
\ In addition to { the} electron filling dependence of $A$ and $J_{o}$, we show the density of states (DOS) of  $\textrm{Co}_2\textrm{MnSi}$ and $\textrm{Co}_2\textrm{MnAl}$ with $\textrm{L}2_{1}$ structure in Fig. \ref{DOS}.
As we can { observe}  from the figure, the Fermi level of $\textrm{Co}_2\textrm{MnSi}$ is located in the gap in the minority spin state and the half-metallic state is achieved.  { Further}, in the DOS of $\textrm{Co}_2\textrm{MnAl}$ with $\textrm{L}2_{1}$ structure, the Fermi level is located near the gap in the minority spin state and the half metallicity becomes worse.
Judging from Fig. \ref{NdepAJ} and Fig. \ref{DOS}, we may conclude that, if the Fermi level is located in the gap,  $A$ and $J_{o}$ become large value because the magnetic excitations caused by spin flip are suppressed due to the absence of the minority spin states at zero temperature. 
\subsection{Finite-temperature properties}
\begin{flushleft} 
\begin{figure}[h]
\begin{center}
\includegraphics[clip,width=9cm]{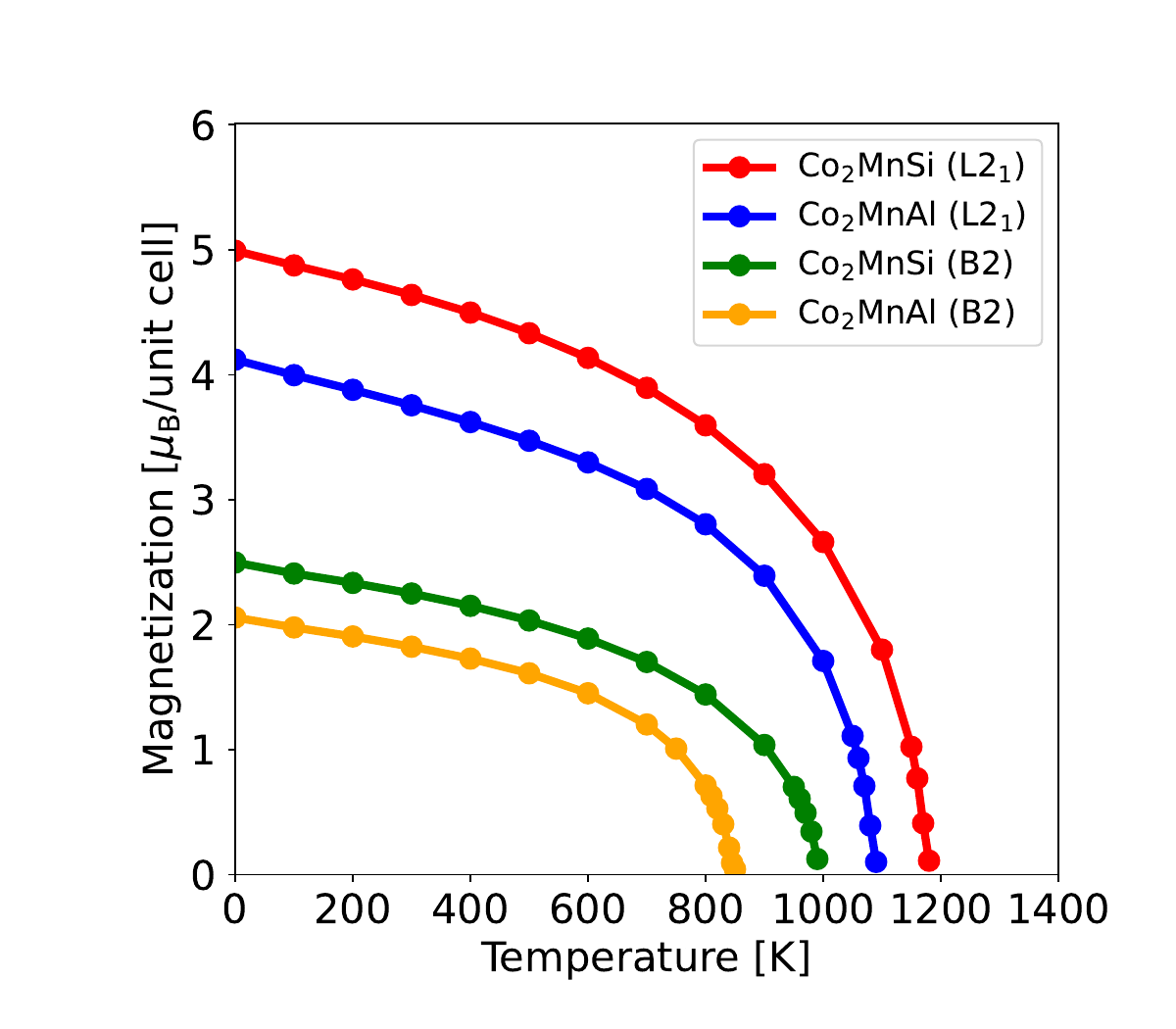}
\end{center}
\caption{Temperature dependence of total magnetization per unit cell for  $\textrm{Co}_2\textrm{MnSi}$ (Red) and $\textrm{Co}_2\textrm{MnAl}$ (Blue) with $\textrm{L}2_{1}$ structures and $\textrm{Co}_2\textrm{MnSi}$ (Green) and $\textrm{Co}_2\textrm{MnAl}$ (Yellow) with $\textrm{B}2$ structures, respectively.}
\label{MT}
\end{figure} 
\end{flushleft}

\begin{flushleft} 
\begin{figure}[h]
\begin{center}
\includegraphics[clip,width=8cm]{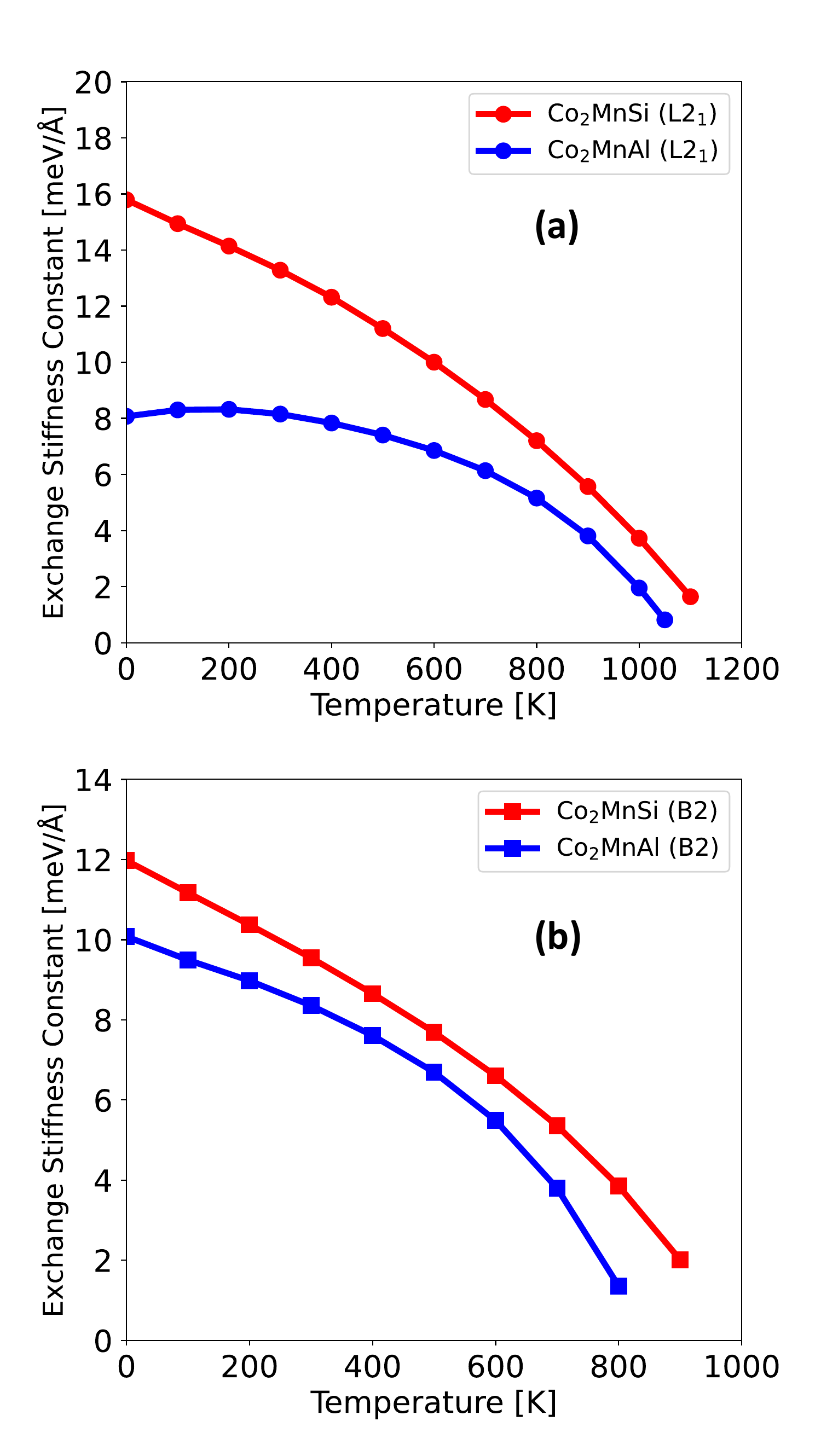}
\end{center}
\caption{Calculated temperature dependence of the exchange stiffness constant of $\textrm{Co}_2\textrm{MnSi}$ and $\textrm{Co}_2\textrm{MnAl}$ with (a) $\textrm{L}2_{1}$ structure and (b) B2 structure.}
\label{AT}
\end{figure} 
\end{flushleft}

\ { To} perform further investigations, we conducted DLM-CPA calculations to examine { the} finite-temperature magnetic properties of these alloys.
We show { the}  calculated temperature dependence of the magnetization of  $\textrm{Co}_2\textrm{MnSi}$ and $\textrm{Co}_2\textrm{MnAl}$ in Fig. \ref {MT}.
{ The} calculated Curie temperatures are 1190  and 1090 K for { the}  $\textrm{L}2_{1}$ structure of $\textrm{Co}_2\textrm{MnSi}$ and $\textrm{Co}_2\textrm{MnAl}$ and 1000 and 850 K for { the} $\textrm{B}2$ structure of $\textrm{Co}_2\textrm{MnSi}$ and $\textrm{Co}_2\textrm{MnAl}$, respectively. In previous works, the Curie temperatures { predicted} with the DLM-CPA method { were}  1103 and 898 K for  { the} $\textrm{L}2_{1}$ and $\textrm{B}2$ structures of  $\textrm{Co}_2\textrm{MnSi}$\cite{Nawa2020} and  1000 and 800 K for { the}  $\textrm{L}2_{1}$ and $\textrm{B}2$ structures of $\textrm{Co}_2\textrm{MnAl}$\cite{Sakuma2022}, respectively. Our results are close to previously calculated values with the DLM-CPA method.
Experimentally, 985\cite{Buscjow1983} and 726 K\cite{Umetsu2008} were reported for $\textrm{Co}_2\textrm{MnSi}$ and $\textrm{Co}_2\textrm{MnAl}$ with $\textrm{L}2_{1}$ structures, respectively. { The overestimation of the} Curie temperature might  be attributed to the single-site approximation in the CPA section.
An astonishing point of our results is that our calculated Curie temperatures { for} $\textrm{Co}_2\textrm{MnSi}$ and $\textrm{Co}_2\textrm{MnAl}$ with { the}  $\textrm{L}2_{1}$ structure have small differences { between them,} 
{although their calculated $A$ and $J_{o}$ values at zero temperature exhibit differences of almost twice between them.}
Here, let us compare { the} results of other type of theories. Chico et al.\cite{Chico2016} calculated the Curie temperatures of $\textrm{Co}_2\textrm{MnSi}$ and $\textrm{Co}_2\textrm{MnAl}$ by using the effective Heisenberg Hamiltonian combined with the LAKG formula and the Monte Carlo simulation. They reported 750  and 360 K for $\textrm{Co}_2\textrm{MnSi}$ and $\textrm{Co}_2\textrm{MnAl}$, { respectively, using}  the LSDA and ASA. In addition, they also reported that { the} calculated spin stiffness constant of $\textrm{Co}_2\textrm{MnAl}$ is nearly half that of $\textrm{Co}_2\textrm{MnSi}$. This trend is consistent with their predicted Curie temperatures. They also concluded that using the full potential scheme can improve the results. However, { an even larger}  Curie temperature has been predicted using the ASA in our case of $\textrm{Co}_2\textrm{MnAl}$. 
Furthermore, K\"ubler et al.\cite{Kubler2007,KublerBook} also calculated the Curie temperatures of these alloys with the random phase approximation. Their obtained results are quite in good agreement with experimental results. 
In their method, they estimated { the} exchange energy from spin-spiral excitation energy. 
Judging from our results and their results, mapping to { an} effective Heisenberg model with the LKAG might underestimate the Curie temperature of $\textrm{Co}_2\textrm{MnAl}$, reflecting a small spin stiffness constant at zero temperature.
To examine the origin of { the} high Curie temperature of $\textrm{Co}_2\textrm{MnAl}$ { with the} $\textrm{L}2_{1}$ structure regardless of the small exchange stiffness constant at zero temperature in our method, we calculated the temperature dependence of { the} exchange stiffness constant of  $\textrm{Co}_2\textrm{MnSi}$ and $\textrm{Co}_2\textrm{MnAl}$ with $\textrm{L}2_{1}$ and $\textrm{B}2$ structures.
{ The} calculated results are shown in Fig. \ref{AT}. We can see that { the} calculated temperature dependence of { the} exchange stiffness constant of $\textrm{Co}_2\textrm{MnSi}$ with { the} $\textrm{L}2_{1}$ structure decrease monotonically with increasing temperatures. { Meanwhile}, we can also see that the behavior of the exchange stiffness constant of $\textrm{Co}_2\textrm{MnAl}$ with { the} $\textrm{L}2_{1}$ structure is quite different from that of $\textrm{Co}_2\textrm{MnSi}$. It increases with increasing temperatures up to 200 K, and afterwards it starts to decrease. This means { that the} exchange stiffness constant is robust against increasing temperatures for  $\textrm{Co}_2\textrm{MnAl}$ with { the} $\textrm{L}2_{1}$ structure. If we assume that { the} temperature dependence of the exchange stiffness constant is { the} same as that of $\textrm{Co}_2\textrm{MnSi}$, which is a monotonically decreasing behavior, { a}  lower Curie temperature might be expected for $\textrm{Co}_2\textrm{MnAl}$ and that is consistent with the prediction from exchange interactions at zero temperature. Our results indicate that the Curie temperatures are not simply determined by the exchange interactions at zero temperature.  In contrast to the cases of the $\textrm{L}2_{1}$ structures, the behaviors of { the} exchange stiffness constants of the $\textrm{B}2$ structures at finite temperatures are simple. They decrease monotonically with { increasing} temperatures. 
\ { Experimentally, { the reported} spin  (exchange) stiffness constants $D$ ($A$)  for $\textrm{Co}_2\textrm{MnSi}$ are  $D$\ =\ $466$ meV\ $\textrm{\AA}^2$ \cite{Ritchie2003} and $334$ meV\ $\textrm{\AA}^2$\cite{Umetsu2011} from the temperature dependence of magnetization at low temperature. In the Brillouin light scattering spectroscopy experiment at room temperature, $586$ meV\ $\textrm{\AA}^2$\cite{Hamrle2009,Kubota2009} was reported ($A$ = $14.67$ meV/\AA ).  { Meanwhile}, for $\textrm{Co}_2\textrm{MnAl}$, it { was} reported that $D$ = $190$ meV\ $\textrm{\AA}^2$ ($A$ = $2.996$ meV/\AA )\cite{Kubota2009} from the Brillouin light scattering spectroscopy experiment, and $D$ = $288$ meV\ $\textrm{\AA}^2$ \cite{Umetsu2011} from the temperature dependence of magnetization at low temperature. }
\\
\ As we have seen above, the difference { in} the behaviors of temperature dependence of the exchange stiffness constants of the $\textrm{L}2_{1}$ structures are unique.
{ To} investigate the origin of these behaviors, we calculated { the} temperature dependence of spin polarization at the chemical potential $\mu$ of $\textrm{Co}_2\textrm{MnSi}$ and $\textrm{Co}_2\textrm{MnAl}$ for both structures. The spin polarization ratio $P$ at $\mu$ is defined as follows:
\begin{align}
P(T)=\frac{D^{\uparrow}(T)-D^{\downarrow}(T)}{D^{\uparrow}(T)+D^{\downarrow}(T)}\times 100,
\end{align}
\begin{align}
D^{\sigma}(T)=-\frac{1}{\pi} \textrm{Im} \textrm{Tr}_{iL} \int \textrm{d} {\bf e}_{i} \  \omega_{i}({\bf{e}}_{i},T) G^{\sigma\sigma}_{ii}(\mu,{\bf{e}}_{i}),
\end{align}
where $\mu$ is determined so that { the} following condition is satisfied:
\begin{align}
&N_{\textrm{e}}=\nonumber \\
&-\frac{1}{\pi} \textrm{Im} \textrm{Tr}_{iL\sigma} \int_{-\infty}^{\infty} \textrm{d} \epsilon f(\epsilon,T,\mu)  \int \textrm{d} {\bf e}_{i} \  \omega_{i}({\bf{e}}_{i},T) G^{\sigma\sigma}_{ii}(\epsilon,{\bf{e}}_{i}).
\end{align}
{ From these figures, we can see that the spin polarization $P(T)$ of $\textrm{Co}_2\textrm{MnSi}$ monotonically decreases with { increasing} temperatures for both structures. These results are consistent with the results calculated by Nawa et al.\cite{Nawa2020}. This clearly indicates that the half metallicity is broken by the thermal spin fluctuations in $\textrm{Co}_2\textrm{MnSi}$. { Meanwhile}, the spin polarization of $\textrm{Co}_2\textrm{MnAl}$ increases with { growing} temperatures at { the} low temperature region for { the} $\textrm{L}2_{1}$ structure. This means that the half metallicity is effectively induced due to the spin disorder in $\textrm{Co}_2\textrm{MnAl}$ with { the} $\textrm{L}2_{1}$ structure. In addition, we can see that { the} temperature dependence of $P(T)$ of $\textrm{Co}_2\textrm{MnAl}$ with { the} $\textrm{B}2$ structure is more robust { than that} of $\textrm{Co}_2\textrm{MnSi}$.\\
\ It is worth mentioning { the} following fact. Oogane et al.\cite{Oogane2006} experimentally measured { the} temperature dependence of the TMR ratios for $\textrm{L}2_{1}$ $\textrm{Co}_2\textrm{MnSi}$- and $\textrm{B}2$ $\textrm{Co}_2\textrm{MnAl}$-based MTJs. They showed that the temperature dependence of { the} TMR ratio for { the} $\textrm{B}2$ $\textrm{Co}_2\textrm{MnAl}$-based MTJ is more robust than the MTJ based on { the} $\textrm{L}2_{1}$ $\textrm{Co}_2\textrm{MnSi}$. { These experimental facts may be connected to our temperature dependence of { the} spin-polarization ratio of these alloys obtained by the DLM picture. }
 \begin{flushleft} 
\begin{figure}[h]
\begin{center}
\includegraphics[clip,width=8.5cm]{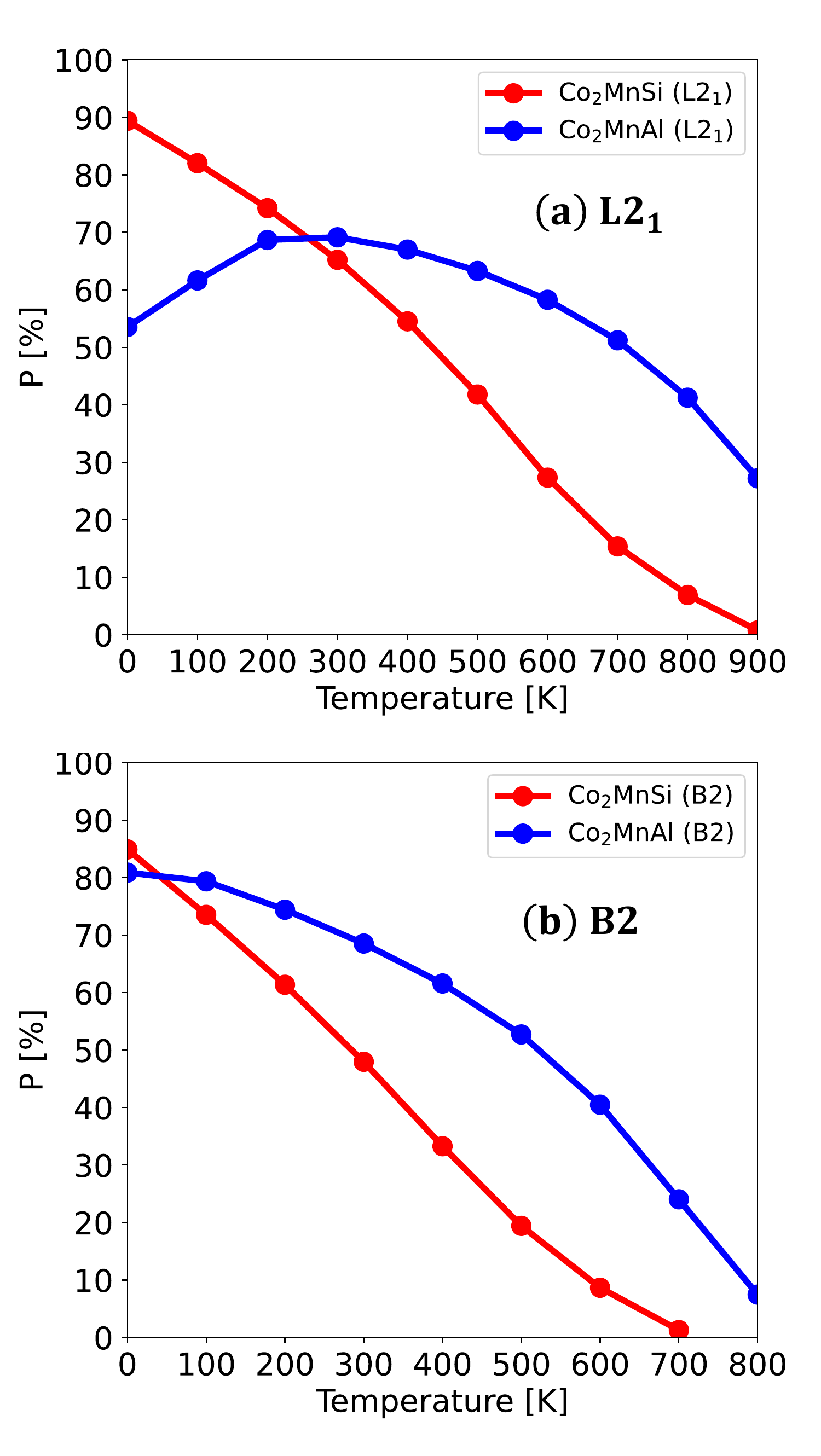}
\end{center}
\caption{Temperature dependence of spin polarization at the chemical potential $\mu(T)$ of $\textrm{Co}_2\textrm{MnSi}$ and $\textrm{Co}_2\textrm{MnAl}$ with (a) $\textrm{L}2_{1}$ structure and (b) $\textrm{B}2$ structure.  }
\label{PCMAS}
\end{figure} 
\end{flushleft}
Finally, { to} confirm what we discussed above, we examined the temperature dependence of the  density of states of these alloys near the chemical potential, focusing on $\textrm{L}2_{1}$ structures.
{ The} calculated results are shown in Fig. \ref{CMAS_DOS2}. From Fig. \ref{CMAS_DOS2} (a), we can { observe} that the gap in the minority spin states starts to collapse and the states start to be occupied with { increasing} temperatures. This means that { the} half metallicity of $\textrm{Co}_2\textrm{MnSi}$ is monotonically destroyed by the spin disorder due to the thermal spin transverse fluctuations. In the case of $\textrm{Co}_2\textrm{MnAl}$, { shown} in Fig. \ref{CMAS_DOS2} (b), although the gap in the minority spin states itself  is destroyed by the spin disorder, the states in the minority spin state become less occupied with increasing temperature in the range up to 400 K. This means that { the} half metallicity is recovered effectively by the thermal spin transverse fluctuations in $\textrm{Co}_2\textrm{MnAl}$. This trend is consistent with the behavior of the temperature dependence of the spin polarization ratio discussed above and has been also confirmed by Sakuma et al.\cite{Sakuma2022} for the temperature dependence of { the} spin-resolved longitudinal conductivity { with the} $\textrm{Co}_2\textrm{MnAl}$ of $\textrm{L}2_{1}$ structure. 
Judging from the results in Fig. \ref{NdepAJ},  Fig. \ref{AT}, and Fig. \ref{CMAS_DOS2}, the spin disorder effect due to thermal spin fluctuations effectively induces half metallicity and it may lead to the change of exchange interactions. { Further,} this may contribute to the robustness of the exchange stiffness constant at finite temperatures and lead to high Curie temperature, in contrast to { the} small exchange stiffness constant at zero temperature. { In addition, this effect might be connected to the spin configuration dependence of the exchange interactions in metallic magnetic systems, as suggested by several authors.\cite{Small1984,Heine1990,Luchini1991,Mryasov1996,Khmelevskyi2007}. 
 In our case, it could be considered that it is caused by the renomarization of the electronic structure due to the scattering potential of the spin disorders. }
\begin{flushleft} 
\begin{figure}[h]
\begin{center}
\includegraphics[clip,width=8.5cm]{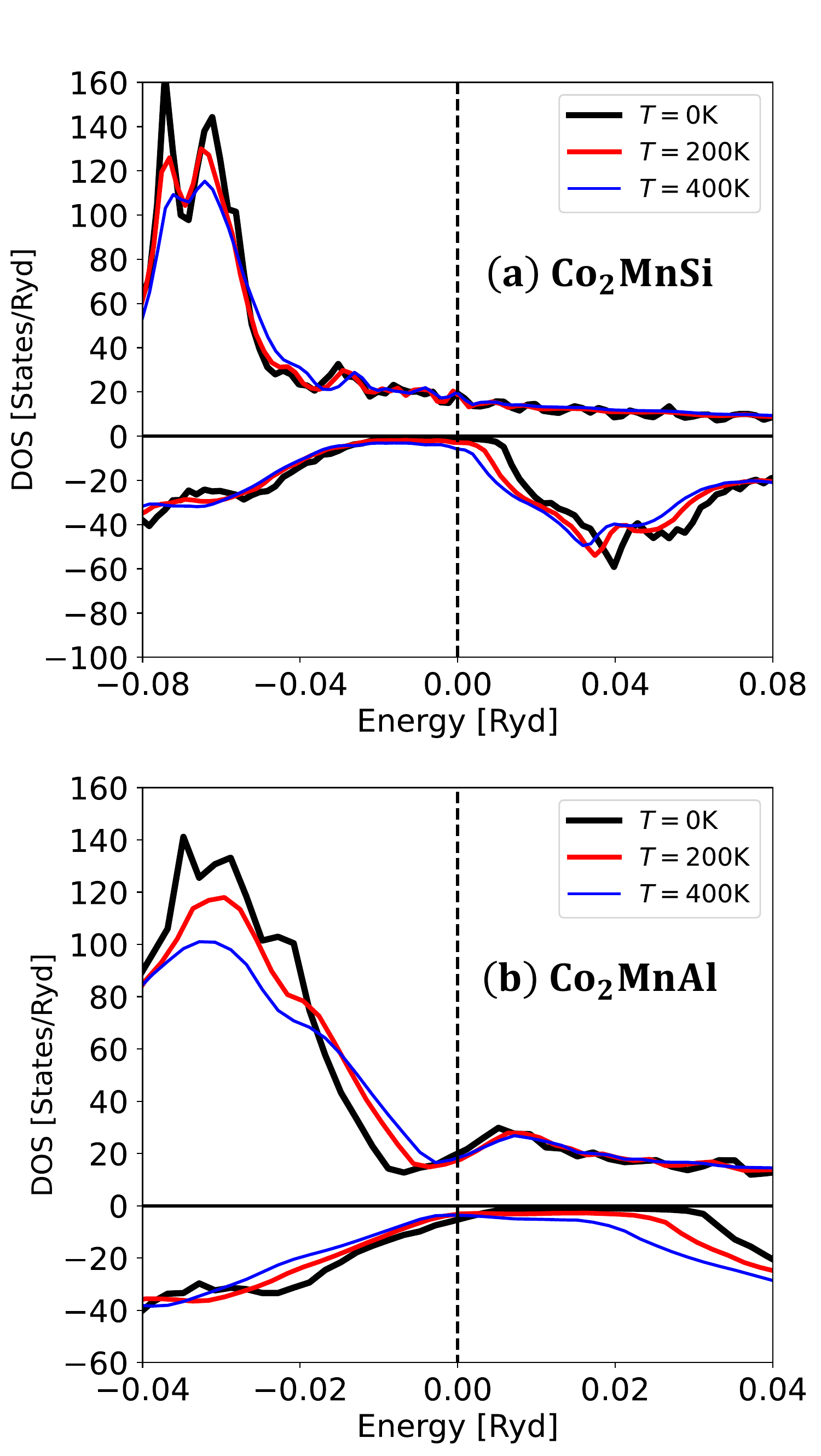}
\end{center}
\caption{Temperature dependence of the { density of states (DOS)} of (a) $\textrm{Co}_2\textrm{MnSi}$ and (b) $\textrm{Co}_2\textrm{MnAl}$ with $\textrm{L}2_{1}$ structure. { The} vertical lines correspond to the chemical potential for each temperature. }
\label{CMAS_DOS2}
\end{figure} 
\end{flushleft}
\ However, several points should be improved { in this study}.
Our finite-temperature treatment neglects the longitudinal spin fluctuation. However, the importance of the longitudinal fluctuation for exchange interactions of Co atoms in $\textrm{Co}_2\textrm{FeSi}$\cite{Khmelevskyi2022} { has been recently indicated}. We also confirmed that the magnetic moments of Co atoms vanish in the self-consistent DLM calculations, which include only up and down spin states with equal ratio in the CPA section, for both $\textrm{Co}_2\textrm{MnSi}$ and $\textrm{Co}_2\textrm{MnAl}$ with $\textrm{L}2_{1}$ structures. { This means { that} $d$ electrons in Co atoms are more itinerant than those in Mn atoms, and their magnetic states might be close to { those in} fcc Co\cite{Uhl1996,Rosengaard1997,Halilov1998}.} In addition, it might be considered that the magnetic moments of Co atoms are sustained by the longitudinal spin fluctuations near Curie temperatures or above\cite{Moriyabook,Hasegawa3,Heine1981}. Our treatment is based on the force theorem, and { the} variation { in} length of magnetic moments is neglected. This might be problematic near Curie temperatures. Therefore, the effect of longitudinal fluctuations for the half metallicity\cite{Santratskii2008} of  $\textrm{Co}_2\textrm{MnSi}$ and $\textrm{Co}_2\textrm{MnAl}$ should be investigated in { a} further study. 
However, { the} unique behavior of the electronic structure of $\textrm{Co}_2\textrm{MnAl}$ is found in { the} low temperature region, and thus, our approximation might not be problematic for these phenomena. \\
\ Our treatment is based on static approximations. According to the results with the dynamical mean field theory (DMFT)\cite{Chioncel2008}, which can include { the} dynamical component of spin fluctuations,  it has been shown that { the} non-quasiparticle states described by the local self energy contribute to { a} reduction { in the}  spin polarization ratio of $\textrm{Co}_2\textrm{MnSi}$ at finite temperatures. This effect on the result of the DOS of $\textrm{Co}_2\textrm{MnAl}$ at finite temperatures could be a target of { research} from the theoretical view.  { However, it may be worth mentioning that non-quasi particles states near Fermi level at finite temperatures and altering the DOS in the majority spin state, both predicted from DMFT calculation results, are not confirmed experimentally in $\textrm{Co}_2\textrm{MnSi}$\cite{Miyamoto2009}.} Furthermore, { investigating} beyond the single-site approximation in the CPA section { could also}  be of interest { for}  further { studies}.
\section{Summary}
In this study, we investigated { the} magnetic properties, such as exchange stiffness constants, not only { at zero temperature} but also at finite temperatures, of Co-based full-Heusler alloys, $\textrm{Co}_2\textrm{MnSi}$ and $\textrm{Co}_2\textrm{MnAl}$, assuming $\textrm{L}2_{1}$  and $\textrm{B}2$ structures  with the DLM-CPA method. 
We confirmed { the} relatively small exchange stiffness constant for $\textrm{Co}_2\textrm{MnAl}$ with { the} $\textrm{L}2_{1}$ structure at zero temperature compared to that of $\textrm{Co}_2\textrm{MnSi}$ with { the} $\textrm{L}2_{1}$ structure. 
This might be related to { the} half-metallic electronic structures and { the} filling dependence of exchange interactions. However, { the} calculated Curie temperature of $\textrm{Co}_2\textrm{MnAl}$ with { the} $\textrm{L}2_{1}$ structure is not { very} different from that of $\textrm{Co}_2\textrm{MnSi}$ with $\textrm{L}2_{1}$ structure, contrary to the expectation from the exchange stiffness constant at zero temperature.
The behavior of { the} calculated temperature dependences of the exchange stiffness constant of $\textrm{Co}_2\textrm{MnAl}$ with { the} $\textrm{L}2_{1}$ structure is unique compared to that of $\textrm{Co}_2\textrm{MnSi}$ with { the} $\textrm{L}2_{1}$ structure. It shows robust behavior against { increasing} temperatures in { the} low temperature region. 
This behavior of the exchange stiffness constant might contribute to { determine the high} Curie temperature of $\textrm{Co}_2\textrm{MnAl}$ with { the} $\textrm{L}2_{1}$ structure. { Meanwhile}, { regarding} the results of the $\textrm{B}2$ structures, the exchange stiffness constants monotonically decrease with { increasing} temperatures for both alloys.  \\
\ To investigate the origin of the behavior of the exchange stiffness constant of $\textrm{Co}_2\textrm{MnAl}$ with { the} $\textrm{L}2_{1}$ structure at finite temperatures, we examined the temperature dependence of the spin polarization ratio at  $\mu$ of both alloys. 
For $\textrm{Co}_2\textrm{MnSi}$, the gap in the minority spin states is destroyed due to the thermal spin disorder, and it leads to { a} monotonically decreasing behavior of { the} spin polarization in both structures. However, in the case of $\textrm{Co}_2\textrm{MnAl}$ with { the} $\textrm{L}2_{1}$ structure, although the gap in the minority spin states is also destroyed, the population of electrons in the minority spin states at $\mu$ { is reduced} due to the change of the DOS caused by spin disorder. It might be considered that this leads { to} the increase { in} spin polarization at { the} low temperature region and the robustness of the exchange stiffness constant against { increasing} temperatures. In addition, the temperature dependence of the spin polarization at $\mu$ of  $\textrm{Co}_2\textrm{MnAl}$ with { the} $\textrm{B}2$ structure { is also} robust with { increasing} temperature compared with { that of} $\textrm{Co}_2\textrm{MnSi}$. \\ 
\ Our result may indicate that the Curie temperatures of itinerant magnets are not simply determined by the exchange interactions at zero temperature, and { the} temperature variations of { the} electronic structure due to spin disorders are also important for the exchange interactions at finite temperatures of the systems.
\begin{acknowledgments}
This work was partly supported by the Center for Science and Innovation in Spintronics (CSIS) and the Center for Innovative Integrated Electronic System (CIES) in Tohoku University and the SIP project, and the X-nics project.
\end{acknowledgments}

\end{document}